# Perpendicular magnetic anisotropy and half metallicity in $Cr_2Ge_2Te_6$ nanoribbons

Valeria Ríos-Vargas, Rodrigo Ponce-Pérez, María G. Moreno-Armenta, and Jonathan Guerrero-Sánchez

*Centro de Nanociencias y Nanotecnología, Universidad Nacional Autónoma de México, km.107, Apdo. Postal 14. Carretera Tijuana-Ensenada, Ensenada, Baja California, México.*

**Abstract**

The $Cr_2Ge_2Te_6$ (CGT) compound is an intrinsic ferromagnetic material with a van der Waals layered structure that shows great promise in spintronics applications. In this work, we investigated the edge effect in the formation of CGT nanoribbons with different terminations as well as the change in electronic and magnetic properties considering spin-orbit effects on the calculations. We studied the thermodynamic stability of the nanoribbon employing the surface formation energy formalism. According to the calculations, in regions of Ge-rich, Te-rich, and Ge-poor, Te-poor of chemical potential the nanoribbon with TeCr edges is the most stable structure, while in the regions of Ge-rich, Te-poor and Ge-poor, Te-poor the nanoribbon with TeCr edges and Te vacancies is the most stable structure. Furthermore, calculations show that the nanoribbon is ferromagnetic with half metal characteristics.

**Introduction**

Magnetic materials have been used until today as key parts of spintronics devices. However, these technologies are approaching a limit in which physical factors preclude their scalability due to the constant miniaturization of devices (Peercy, 2000). Either discovery or engineering of materials with tuned properties is needed to construct devices without these limitations, a goal that is still evident in the scientific community.

The discovery of perpendicular magnetic anisotropy (PMA) has brought a way to break such physical limitations since storage is now out-of-plane and through the stacking of magnetic tunnel junctions (MTJs) in several architectures, improving the storage capacity while reducing the device's size and increasing the density (Levi, 2004). Magnetic random access memories take advantage of PMA and are expected to become the future of random access memories (Sbiaa et al., 2011). An MTJ is commonly constructed by a pinned layer, an insulating barrier, and a free layer, which will generate the magnetoresistance effect and further information storage (Tudu & Tiwari, 2017). Several materials have been treated as free and pinned layers, being CoFeB (Ikeda et al., 2010), MnGa (Mao et al., 2017), and some Heusler alloys (Kabanov et al., 2021) the primary candidates. MgO is usually the best choice as an insulating layer since it provides high spin selectivity and large tunnel magnetoresistance (Beletskii et al., 2007).

A natural way to eliminate the scalability issue is to consider the recently discovered intrinsic 2D magnetic materials like CGT (Gong et al., 2017) and CrI3 (Huang et al., 2017), which possess large spin polarization and intrinsic PMA. Their integration in MTJs is key to produce atomic-scale devices. 2D tunnel barriers may also allow effective thickness control. In this way, h-BN and graphene have been effectively used as single-layer tunnel barriers (Piquemal-Banci et al., 2020). The main idea here is to produce few-layer MTJs, which will naturally drive atomic-scale devices with enhanced storage capacity, high PMA, flexibility, reduced heating, and low power consumption (Zhang et al., 2021).

2D materials as part of MTJs are the first attempts to eliminate some drawbacks presented by 3D MTJs. The interface quality, together with interface bonding between magnetic layers and the insulating barrier, are major concerns. These issues affect the magnetoresistance in the devices (Wang et al., 2018). Integrating a 2D insulating barrier would help since a van der Waals gap appears, allowing enhanced spin transfer from the fixed layer to the free layer (Wang et al., 2018). Several 2D materials like $CrI_3$, $Fe_3GeTe_2$, $MoS_2$, Graphene, h-BN, etc, have been integrated into 2D MTJs (Zhang et al., 2021). However, this is an active research area in which new 2D materials need to be developed to achieve enhanced spin filtering. One of the main conditions to improve spin filtering is to have effective control of the spin polarization. In this way, half-metallic materials are highly desired. A half-metal material depicts a hundred percent spin polarization because one spin channel behaves as metal while the other presents a semiconductor behavior.

On the other hand, CGT is an intrinsic 2D long-range ferromagnetic sheet with an out-of-plane easy axis, exhibiting an apparent PMA effect (Gong et al., 2017). Its physical properties are easily tuned by electrostatic grating, modifying its magnetic phase transition and magnetic anisotropy (MAE) (Verzhbitskiy et al., 2020). An efficient spin-orbit-torque switching effect has been observed in CGT interfaced with Ta and Pt (Ostwal et al., 2020, Gupta et al., 2020). Also, the curie temperature and MAE can be greatly enhanced by interfacing CGT with $PtSe_2$, in which Dzyaloshinskii-Moriya interactions and single-ion anisotropy dominate such behavior (Dong et al., 2020). MAE is also enhanced in CGT by an induced pressure (Sakurai et al., 2021). Furthermore, a robust half-metal effect emerges in CGT monolayer by electron doping and adsorption of alkali metals (Ilyas et al., 2021). As previously discussed, CGT is a magnetic 2D material with very tunable properties; then, it is cataloged as key in the further development of few-layer spintronics devices.

Since the integration of half metal materials in MTJs is an effective way to improve the magnetoresistance and spin-transfer. We propose the existence of a half-metal effect on CGT nanoribbons, which naturally appears when cutting the 2D sheet in the most stable terminations. Also, half-metallicity emerges by defect engineering when removing Te atoms from the TeCr edge terminated ribbon. Such an exciting and applicable effect emerges because of the formation of chain-like structures at the edges. The Cr-d contributions are the ones that induce the metallic behavior in the majority carriers. In contrast, the Te-p orbitals dominate the highest occupied orbital of the minority carrier's channel. The 100% spin polarization found in CGT ribbons points to this material as a potential pinned layer of an MTJ.

**Computational Methodology**

To study CGT in its bulk, monolayer, and nanoribbon form, first-principles calculations based on density functional theory (DFT) were implemented using the PWscf code of the Quantum ESPRESSO package with projected augmented wave (PAW) pseudopotentials (Blöchl, 1994). In addition, the exchange-correlation energies were treated according to the generalized gradient approximation (GGA) (Perdew et al., 1996). The van der Waals interactions were incorporated using the Grimme DFT-D3 method (Grimme et al., 2010). To investigate the electronic and magnetic properties, spin-polarized calculations and a k-points mesh of 9 x 9 x 3 have been used. Optimized cut-off energy of 42 Ry was used for all calculations. The charge density cutoff followed the Ecut-rho=8Ecut-off rule. To investigate the magnetic anisotropy energy, we used the Vienna Ab initio Simulation Package (VASP) with an energy cut-off of 450 eV. Similar conditions were set for the remaining parameters

## Results and Discussion

### A. Structural properties of CGT in bulk and monolayer

Calculations were performed on the CGT in its bulk and monolayer form to build the nanoribbons. The bulk material presents a hexagonal crystalline structure with optimized lattice parameters in *a = b = 7.00 Å* and in *c = 20.60 Å*, which agrees with the experimental values of *a = b = 6.83 Å* and *c = 20.56 Å* (Hao et al., 2018). **Figure 1a** shows the CGT bulk structure. It is observed that this structure presents a layered structure with strong in-plane chemical bonds whereas out-of-plane, the layers are joined together by weak van der Waals (vdW) interactions.

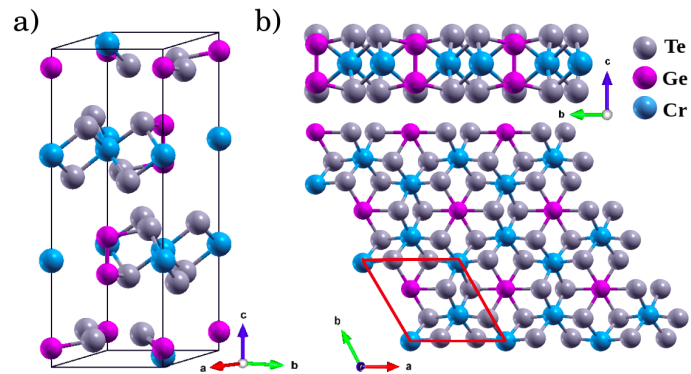

**Figure 1**. The CGT crystal structure in its a) bulk and b) monolayer form. Gray, magenta, and blue stand for Te, Ge, and Cr atoms. The unit cell of the monolayer is shown in red in b).

On the other hand, the monolayer presents an optimized lattice parameter *a=6.9 Å*, concordant with reported values of 6.857 Å and 6.8275 Å, by Kangying Wang, et al.*l* in 2019 and by A.K. Nair, *et al.* in 2020, respectively. Both bulk and monolayer structures have a space group $R\bar{3}$ consisting of a honeycomb-shaped lattice formed by Cr atoms as shown in **Figure 1**, each Cr is contained in the center of a Te octahedron. Note also the formation of Ge dimers, see **Figure 1b**.

Once the optimized CGT monolayer, we construct the nanoribbons in a 3x3 supercell with an empty space in the z-direction of 10 Å. The new optimized lattice parameter is a=20.4 Å. The defined edge terminations are as follows: TeGe-terminated nanoribbons (both edges), TeCr edges, TeCr termination on the left edge and TeGe on the right edge, TeGe edges on the left side, and TeCr edges on the right side, both edges TeCr with Te vacancies on both edges, terminations of TeGe on the left edge and TeCr on the right edge with tee vacancies on both edges, TeGe on the left edge with Ge vacancies and TeCr on the right edge with Cr vacancies, and finally the wide nanoribbon built into a 3x4 supercell with both edge terminations from TeGe.

## B. Thermodynamic stability

Eight possible atomic arrangements with different numbers of atoms emerge at the edges of the nanoribbons. When systems do not possess the same number of atoms, the total energy is not a stability criterion to discern between models. Instead, a formalism independent of the number of atoms is needed. Formation energy depends on the chemical potentials of the involved species, and indeed it helps to clarify the stability of models with a different number of atoms. Here, we compare the formation energy of various edges to determine the most favorable to appear in the experiment. Likewise, each nanostructure was optimized, resulting in an optimal value in b of 20.46 Å. The studied nanoribbons are shown in **Figure 2**. We treated nanoribbons with TeGe edges (**Figure 2a**) and TeCr edges (**Figure 2b**). Nanoribbons with TeGe edge in the left and TeCr edge in the right (**Figure 2c**), TeCr edge in the left and TeGe edge in the right (**Figure 2d**), with TeGe edges in a larger 4x3 periodicity (**Figure 2e**). A nanoribbon with Te vacancies in TeCr edges (**Figure 2f**), TeGe in the left edge and TeCr in the right edge, Te vacancies in both edges (**Figure 2g**), and a TeGe terminated nanoribbon with Ge vacancies in the left edge, the right edge is TeCr terminated with Cr vacancies (**Figure 2h**). The red, yellow, and green circles in the Figure exemplify the different vacancies possible in the nanoribbons. For example, if we remove the Te atoms at the edge of Figure 2b, we achieve the nanoribbon in Figure 2f. After structural relaxation the nanoribbon rearranges to form Ge-Ge, Cr-Ge, Te-Ge, and Cr-Te bonds. Also, by removing the Te atoms in the nanoribbon of Figure 2c drives to the nanoribbon in Figure 2g. The Te vacancies generate Ge-Ge dimmers at the left edge whereas at the right, they form Ge-Ge, Ge-Cr, Cr-Te and Te-Ge bonds. Finally, upon removing Ge atoms from the left edge and Cr atoms from the right edge of the nanoribbon in Figure 2c, we achieve the nanoribbon in Figure 2h, such nanoribbon shows only Te atoms at both edges.

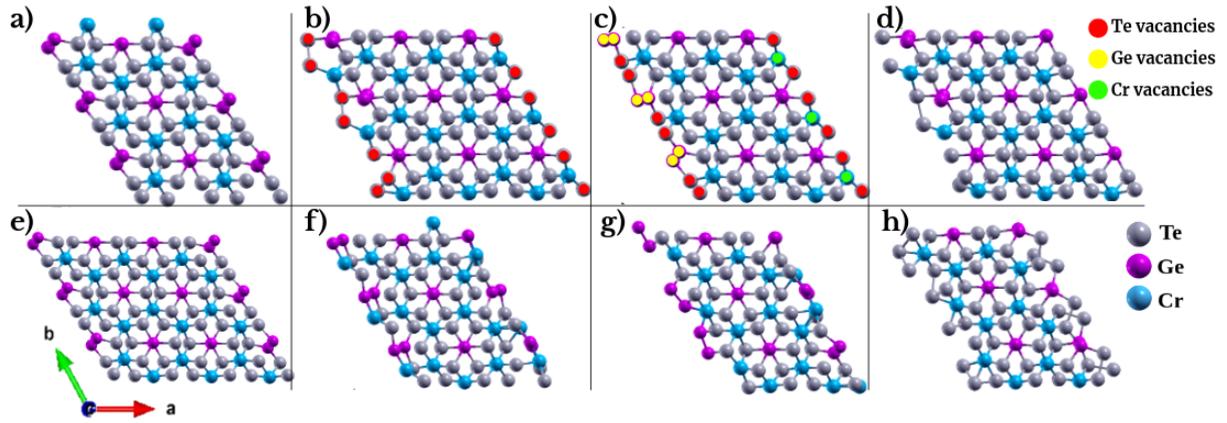

**Figure 2**. CGT nanoribbons with a) TeGe edges, b) TeCr edges, c) TeGe on the left edge and TeCr on the edge right, d) TeCr on the left edge and TeGe on the right edge, e) TeGe edges in a 3x4 supercell, f) Te vacancies in TeCr edges, g) TeGe on the left edge and TeCr on the right edge, Te vacancies in both edges, and h) TeGe on the left edge with Ge vacancy, TeCr on the right edge with Cr vacancy.

The nanoribbons thermodynamic stability was obtained using the edge formation energy formalism (Bollinger et al., 2003). The following expression was defined:

$$\gamma = \frac{1}{A} [G_{nanoribbon} - \frac{n_{Cr}}{2}\mu_{Cr2Ge2Te6} + \mu_{Ge}(n_{Cr} - n_{Ge}) + \mu_{Te}(3n_{Cr} - n_{Te})], \quad (1)$$

where $A$ is the nanoribbon area, $G_{nanoribbon}$ is the Gibbs free energy of the nanoribbon, $n_i$ and $\mu_i$ are the numbers of atoms inside the nanoribbon, and the chemical potential, respectively. The most stable nanoribbon will be the one that minimizes $\gamma$. The edge free energy equation allows us to determine the ranges of chemical potentials in which the structures exist with the most stable edge configuration. Under this approximation, the Gibbs free energy of the nanoribbon ($G_{nanoribbon}$) corresponds to the DFT energies from nanoribbon configuration. The nanoribbon formation energies shown in Figure 3 reveal that the most stable nanoribbon is the one with TeCr edges for Ge-rich, Te-rich, and Ge-poor, Te-rich conditions and intermediate growth conditions (see Figure 2b for structural details). On the other hand, Te vacancies stabilize the TeCr edges for a reduced range of chemical potentials that correspond to Ge-rich, Te-poor, and Ge-poor, Te-poor conditions (see Figure 2f for structural details). Table I summarizes the formation energies of the most stable nanoribbons at different growth conditions. Notice from the stability plots that the remaining edge models do not achieve stability for any range of chemical potential. Te atoms are not stable at the edge but it can achieve stability if they form bonds with Cr atoms.

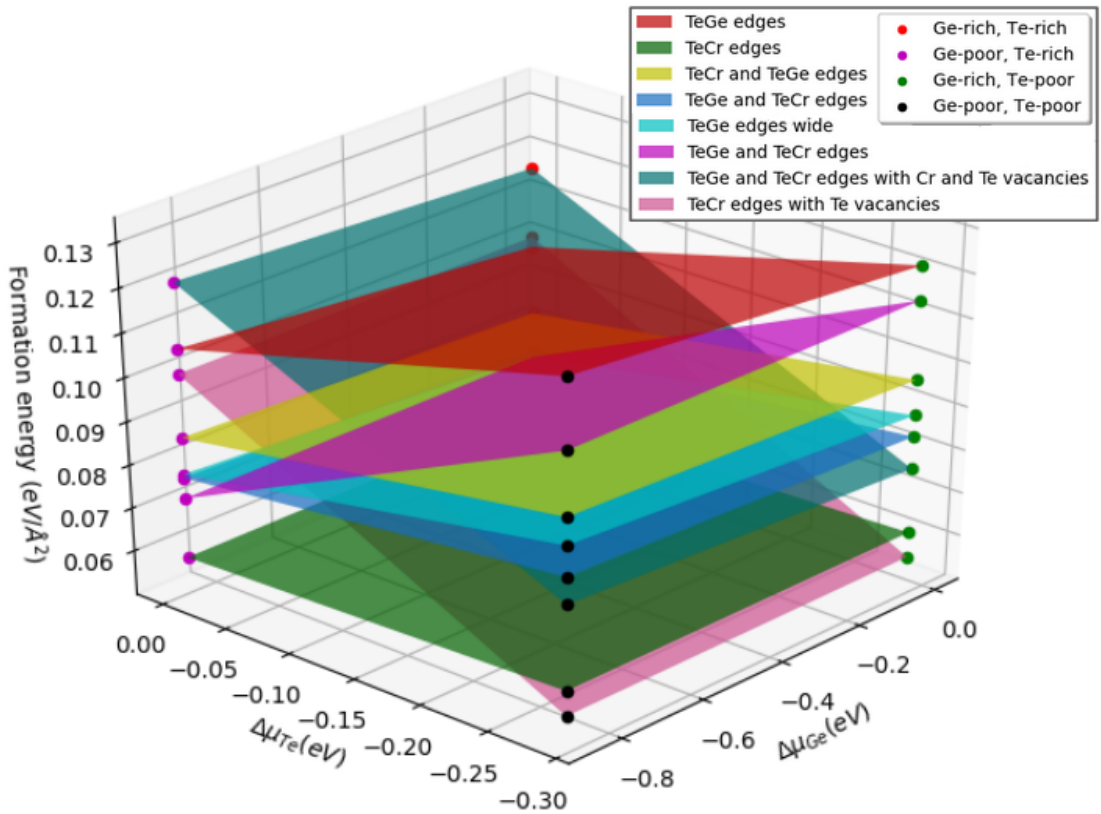

**Figure 3.** Edge formation energy for nanoribbons with different edges. A color code is used to define each nanoribbon. Circles of different colors are to highlight the growth conditions. Red circle defines the Ge-rich, Te-rich conditions, magenta establishes the Ge-poor, Te-rich conditions, green the Ge-rich, Te-poor conditions, and black defines the Ge-poor, Te-poor growth conditions. Different views of the edge formation energy plot are available in the Supplementary section.

Table I. Formation energies of the most stable nanoribbons at different growth conditions.

| Growth conditions | | Edge Formation Energies (eV) | |
|---|---|---|---|
| | | TeCr edges | TeCr edges with Te vacancies |
| Ge-rich | Te-rich | 0.058 | 0.101 |
| Ge-poor | Te-rich | 0.058 | 0.101 |
| Ge-rich | Te-poor | 0.061 | 0.055 |
| Ge-poor | Te-poor | 0.061 | 0.055 |

## B. Structural properties

Once the most stable nanoribbons were defined, we perform a structural analysis to observe the change in the bond lengths present at the edges. Likewise, the monolayer bonds (Table II) were compared to the existing bonds in each nanoribbon. **Figures 4a** and **4b** show the nanoribbons structures with TeCr edges and Te vacancies in TeCr edges, respectively. Edge regions are highlighted; the left region (LR) in blue, and the right region (RR) in green. Edge effects generate Te-Te, Cr-Cr, and Cr-Ge bonds in LR and RR that are not present in the monolayer (**See Figures 4a and 4b**). In the LR region of the TeCr edge, Te-Te bonds are generated with an average length of 2.812 Å. On the other hand, the nanoribbon with Te vacancies at the edge presents Cr-Cr bonds with an average distance of 2.887 Å in LR and Cr-Ge bonds with an average length of 2.515 Å in RR.

Table II. Bonds present in the CGT monolayer.

| Bonds | Length (Å) |
|---|---|
| Te-Ge | 2.62 |
| Te-Cr | 2.78 |
| Ge-Ge | 2.41 |

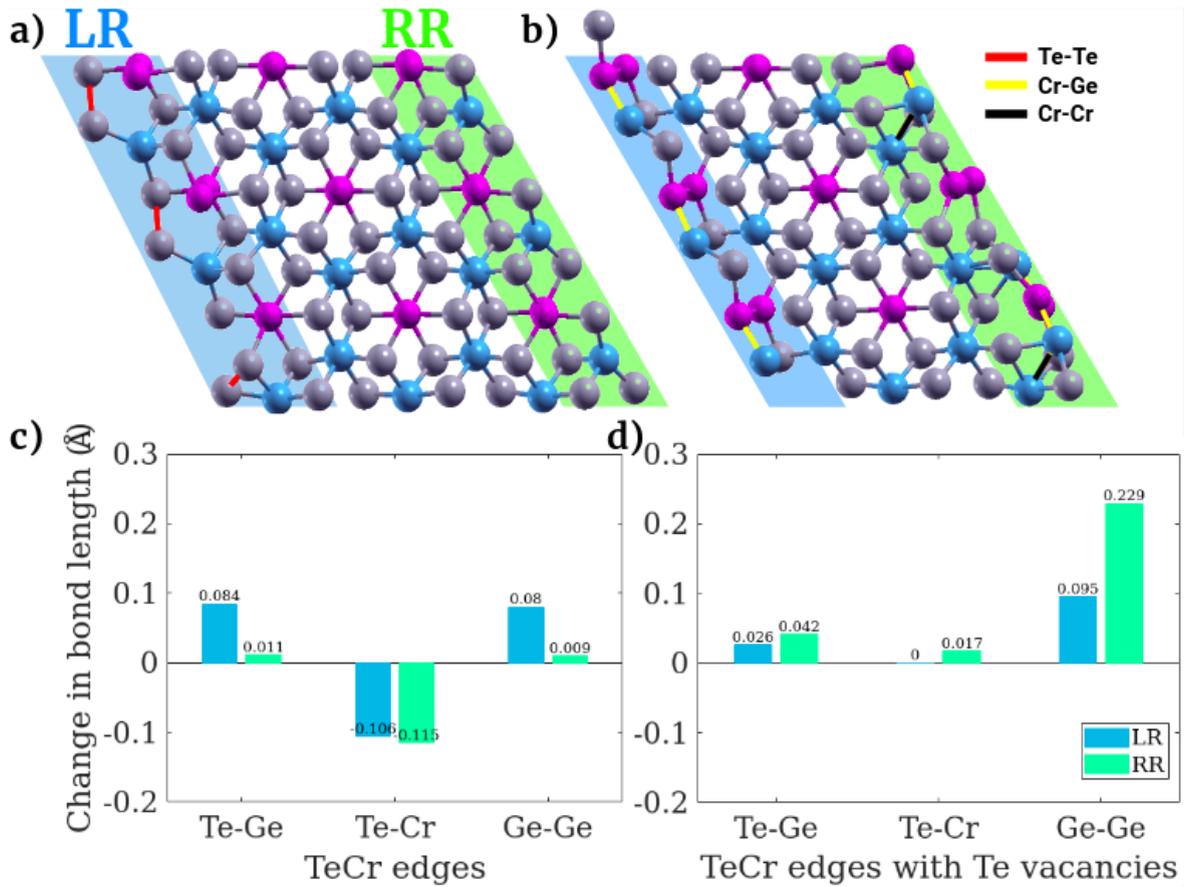

**Figure 4**. Atomic structure of the stable nanoribbons: a) terminated with TeCr edges and b) terminated with TeCr edges, Te vacancies in both edges. Left region (LR) and right region (CR) are highlighted in blue and green, respectively. c) The change in bond lengths at the TeCr edges of the nanoribbon with respect to the monolayer distances, and d) The change in bond lengths at the TeCr edges with Te vacancies as function of the monolayer distances. The Red, yellow, and black lines shown at the edges correspond to the new Te-Te, Cr-Ge, and Cr-Cr bonds appearing due to the nanoribbons formation.

In **Figure 4c** and **4d**, we depict the average bond length of the bonds present in the LR and RR regions of each nanoribbon with respect to the monolayer distances. In the nanoribbon with TeCr edges (**Figure 4c**), it is observed that the Te-Ge and Ge-Ge bonds at the LR edge increase by 0.084 Å and 0.080 Å, respectively, while the Te-Cr bonds decrease their length by 0.106 Å as an edge effect that stabilizes the structure, together with the Te-Te bonds formation. On the other hand, the Te-Ge and Ge-Ge bonds in RR present an almost negligible change of 0.011 Å and 0.009 Å, respectively. Finally, the Te-Cr bonds decrease 0.115 Å, a similar stabilization effect as in the left side of the nanoribbon that is related to the Te-Cr bonds contraction. In the TeCr nanoribbon edges with Te vacancies (**Figure 4d**), the Te-Ge and Ge-Ge bonds increase in length by 0.026 Å and 0.095 Å, respectively, in LR. In contrast, Te-Cr bonds remain with no change with respect to the monolayer distance. Here the edge stabilized due to the formation of the Cr-Ge bonds. Finally, in RR, the Te-Ge, Te-Cr, and Ge-Ge increase in 0.042 Å, 0.017 Å, and 0.229 Å, respectively. Here we associate the stabilization effect to the Cr-Ge bonds and Cr-Cr dimmers formation.

### C. Electronic and magnetic properties

In this section, we calculated the Density of States (DOS) of the two most stable ribbons by relaxing the structures with Hubbard U=3.9 eV, then we carried out self-consistent calculations (SCF). In **Figure 2a**, the PDOS of the nanoribbon with TeCr edges reveals that the majority spin channel ↑ presents metallic characteristics, while the minority spin channel ↓ behaves like a semiconductor. This behavior is typical of HM materials, so it is concluded that the nanoribbon with TeCr edges exhibits properties of HM materials. In addition, it is observed that the bandgap of the minority channel is shifted to the right. It is observed that the maximum of the valence band (VBM) is dominated by the hybrid of Te-$p$ and orbitals of Cr-$d$, while the conduction band minimum (CBM) is dominated by the hybrid orbital of the

Cr-*d* electrons and Ge-*p*.

In the same way, in the DOS of the nanoribbon with TeCr edges and Te vacancies (**see Figure 2b**), HM characteristics are also observed. However, in this case, the bandgap in the minority channel widens. This broadening arises from the presence of Ge dimers in both edges of the nanoribbon. As in the structure without vacancies, a displacement of the bandgap to the right is observed. Finally, orbitals with the greatest presence in the electrical properties of the nanoribbon are the hybrid Te p orbitals in VBM, while CBM is dominated by the hybrid orbital of Cr-*d* and Ge-*p* electrons.

This result is of great importance for the development of magnetic tunnel junction devices since HM materials are capable of generating polarized spin currents. Because the forbidden gap of the minority channel is wider, in the nanoscale with Te vacancies, there is a greater range of energies in which a polarized spin current will be obtained with respect to the nanoscale without vacancies, so this can be considered the structure of greatest interest for its potential application in spintronic devices.

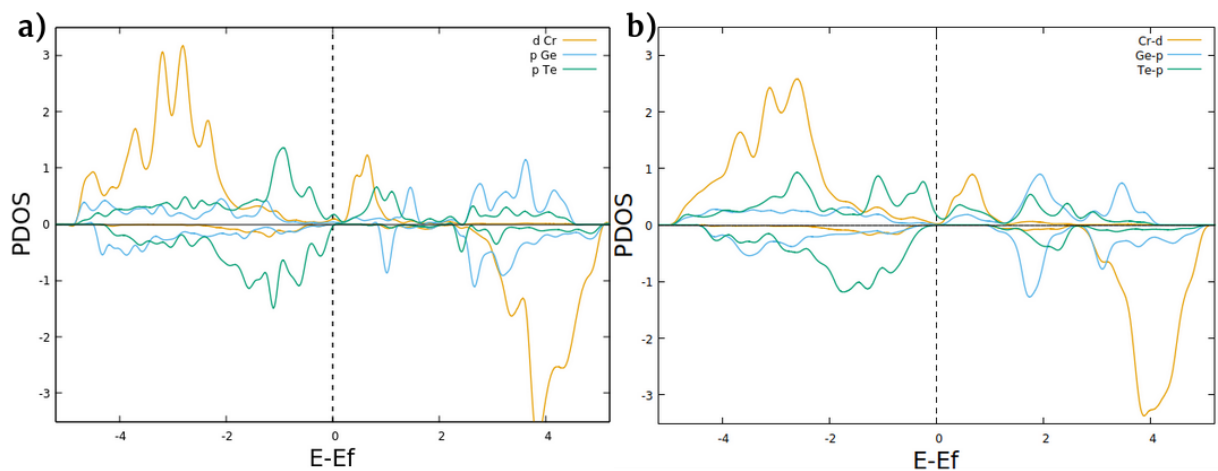

**Figure 5. a)** Density of states and **b)** Projected Density of States of nanoribbon CGT with TeCr edges, c) density of states and d) Projected density of states of the nanoribbon with Te vacancies in the TeCr edges.

To conclude the study of the CGT nanoribbons, the magnetic anisotropy (MAE) analysis was performed. The results show that the energetically favorable direction of spontaneous magnetization in both nanoribbons is out-of-plane, in direction [001]. Table III shows the relative energies of the easy magnetization axis in both nanoribbons.

Table III. Relative energies of the easy magnetization axis with respect to the direction [001]

| Directions | Energy (meV) | |
|---|---|---|
| | TeCr edges with Te vacancies | TeCr edges |
| [0.68 2 0] | 8.10 | 6.50 |
| [0 0 2] | 0.00 | 0.00 |
| [0 2 0] | 7.90 | 10.5 |
| [2 0 0] | 7.70 | 5.50 |
| [1 1 0] | 9.40 | 2.00 |

Subsequently, the atomic magnetic moments present in the nanoribbons and their edges were analyzed and compared with the magnetic moments of the atoms in the monolayer. In Table IV it is observed that the magnetic moments of the Cr atoms have a lower magnitude in the monolayer than in the non-ribbons $0.19\mu B$. In addition, the chromes found on the edges of both nanoribbons increase their magnetic moment by $0.39\mu B$. On the other hand, the magnetic moments of the Ge atoms decrease their magnitude $0.001\mu B$ in the nanoribbons with TeCr edges. However, it is observed that in the nanoscale with Te vacancies the magnetic moments change direction and their magnitude is reduced to $0.01\mu B$, where the magnetic moments have the greatest magnitude of the edge. Finally, the Te atoms present a change in the magnitude of the magnetic moments of $0.01\mu B$, while at the edges their magnitude changes $0.02\mu B$.

**TABLE IV.** Magnetic moments of the atoms present in the monolayer, the nanoribbons, and the edges of each nanoribbon.

| | Magnetic moments ($\mu_B$) |
|---|---|

| Atoms | Layer | TeCr edges | | TeCr edges with Te vacancies | |
|---|---|---|---|---|---|
| | | Ribbon | Edge | Ribbon | Edge |
| Cr | 3.60 | 3.8 | 3.96 | 3.77 | 4.04 |
| Ge | 0.02 | 0.01 | 0.01 | -0.00 | -0.01 |
| Te | -0.17 | -0.17 | -0.15 | -0.17 | -0.14 |

The previous analysis can be complemented with the projection of the spin density in the plane [001]. As shown in **Figure 3**, it is observed that the change in the density of the Cr atoms at the edge with respect to the atoms in the center of the nanoribbons is very small so it cannot be distinguished graphically.

The results obtained from the magnetic properties show that the CGT nanoribbons have properties of great interest for the development of MTJ devices since both structures have perpendicular magnetic anisotropy (PMA). This characteristic is conferred on the material by the edge effects on the material.

**Conclusion**

In the present work, a study of thermodynamic and structural stability was presented, as well as an analysis of the electrical and magnetic properties of CGT nanoribbons. In the study of thermodynamic stability, the edge formation energies formalism was used and it was found that the nanoribbon with TeCr edges is the most stable structure under Ge-rich, Te-rich, and Ge-poor, Te-rich conditions. On the other hand, the nanoribbon with TeCr edges and Te vacancies is stable in Ge-rich, Te-poor, and Ge-poor, Te-poor conditions.

In the structure of the two most stable nanoribbons, new bonds are generated due to edge effects. In the nanoribbon with TeCr edges, Te-Te bonds are generated in the LR region, while in the nanoribbon with TeCr edges and Te vacancies, Cr-Cr bonds appear in the RR region, and Cr-Ge bonds are generated in LR, and RR regions. In addition, in this last structure, the presence of chain-type bonds in LR can be observed.

In the penultimate part of the study, the electronic properties of CGT nanoribbons were analyzed. The CGT material in its nanoribbon form was found to exhibit HM properties. It was possible to conclude that in the nanoribbon with Te vacancies at the edges, the minority channel has a wider prohibited gap of energies compared to the nanoribbon without vacancies. This characteristic is due to the existence of chain-type bonds and Ge dimers present in the edges. This result is very valuable for the development of spintronic devices since HM materials are capable of producing spin-polarized current. Finally, in the study of magnetic properties, it was found that both nanoribbons present perpendicular magnetic anisotropy. In this way, it can be concluded that the nanoribbon with TeCr edges and TeCr edges with Te vacancies is of great interest for the development of devices that use the magnetic tunnel junction principle.